# Core-hole-induced dynamical effects in the x-ray emission spectrum of liquid methanol

M. P. Ljungberg[1], I. Zhovtobriukh[2], O. Takahashi[3], and L. G. M. Pettersson[2]

[1] Donostia International Physics Center, Paseo Manuel de Lardizabal, 4. E-20018 Donostia-San Sebastián, Spain
[2] FYSIKUM, Stockholm University, AlbaNova, S-106 91 Stockholm, Sweden
[3] Institute for Sustainable Sciences and Development, Hiroshima University, 1-3-1, Kagamiyama, Higashi-Hiroshima, 739-8526

**Abstract**

We compute the x-ray emission spectrum (XES) of liquid methanol, with the dynamical effects that result from the creation of the core hole included in a semiclassical way. Our method closely reproduces a fully quantum mechanical description of the dynamical effects for relevant one-dimensional models of the hydrogen-bonded methanol molecules. For the liquid we find excellent agreement with the experimental spectrum, including the large isotope effect in the first split peak. The dynamical effects depend sensitively on the initial structure in terms of the local hydrogen-bonding (H-bonding) character; non-donor molecules contribute mainly to the high-energy peak while molecules with a strong donated H-bond contribute to the peak at lower energy. The spectrum thus reflects the initial structure mediated by the dynamical effects that are, however, seen to be crucial in order to reproduce the intensity distribution of the recently measured spectrum.

# I. Introduction

X-ray spectroscopies, such as x-ray emission (XES) and x-ray absorption spectroscopy (XAS), provide powerful probes of the electronic structure around selected atoms in a molecule or material [1-5]. Due to the involvement of the core-level in the process they give very local information and, based on the relevant selection rules, provide an experimental correspondence to a quantum chemical population analysis of the involved molecular orbitals in a molecular system or the band structure in a material. This is so since for a dipole transition, *e.g.*, from or to a spherically symmetric $1s$ state only states with appreciable local *p*-character can contribute. Furthermore, when the excitation is resonant with some intermediate state of specific symmetry then additional symmetry selection rules apply and provide further information on the system [2,6].

In the present work we will focus on the interpretation of XES of liquid methanol which is one of the simplest hydrogen-bonded (H-bonded) liquids – other examples are the other alcohols and liquid water. These systems share certain features in the oxygen XES spectra, notably a split peak in the region of the molecular HOMO, that is the $2a''$ orbital for methanol, or the $1b_1$ orbital for water. Here there is a debate at present as to how the observed split peak should be interpreted [5, 7-29]. It is clear that core-hole-induced dynamics, or equivalently, life-time vibrational interference effects [30] are important, as evidenced by the large isotope effects in the spectra when substituting the hydrogen atoms for deuterium atoms. [11, 12, 22, 26, 27, 31, 32]. However, there is still an open question about the effect of the instantaneous structure before the core ionization, especially for water that has more possible H-bond arrangements [5, 12, 14, 15, 21, 22, 24-26, 31].

Since methanol only has one hydrogen atom per molecule that can participate in H-bonds (unlike water, which has two) it is a natural starting point in order to test and calibrate theoretical approaches that can later be used in more complicated H-bonded systems. High-resolution experimental XES data for liquid methanol have furthermore recently become available [27].

Simulating the vibrational effects in the XES process one has to take into account the significant zero-point energy that is released as the hydrogen-bonding potential energy surface is strongly modified upon creation of the core-hole. For the short time-scales involved (the O $1s$ core-hole life-time is ~3.6 fs [33]) we have shown that an ensemble of classical trajectories well reproduces the quantum wave packet propagation as long as a sufficient sampling of the initial quantum probability distributions in OH distances and momenta is performed [34]. Using spectra calculated along these trajectories, our semiclassical approximation to the Kramers-Heisenberg formula (SCKH) was shown to give excellent agreement with fully quantum mechanical spectra for a model water dimer [34]. The big advantage of the semiclassical scheme is that an arbitrary number of vibrational modes can be included in the calculation through the classical dynamics, as will be done in the present work.

First of all, we will show that our SCKH method reproduces the full vibrational quantum treatment for one-dimensional problems constructed by considering an H-bonded methanol dimer with bond length varied in the range found in the liquid. We will then apply our approach to reproduce the experimental non-resonant XES spectrum of the liquid for both normal methanol (MeOH) and deuterated methanol (MeOD) [27]. We find that including the dynamical effects or equivalently the vibrational interference effects is crucial for reproducing the experimental spectrum and we show how the intensity of the initially separated molecular peaks redistributes to give the intensity distribution in the split peak feature seen in the experiment. Furthermore, we can connect the dynamical behavior with the local H-bond configuration: strongly H-bonded molecules experience large dynamical effects that in particular contribute to and enhance the low-energy feature in the first split peak, while weakly H-bonded species undergo less dynamics and contribute more to the high-energy peak. In general we obtain very good agreement with the experiment, including the large isotope effect in the first split peak. Our work shows the feasibility of our method for capturing the relevant dynamical effects in realistic models of H-bonded liquids, and we believe that our method will be very suitable to apply to other similar systems.

**II. Methods**

X-ray emission spectra can be well described by the Kramers-Heisenberg (KH) formula [2], which is a one-step description of the absorption of an incoming photon, creating a core-excited state, and the emission of an outgoing photon where an electron falls down to occupy the core orbital. The full one-step process is also denoted resonant inelastic x-ray scattering (RIXS) [2]. In some cases these two processes can be separated, but this neglects interference between intermediate states. Such interference effects turn out to be crucial in order to describe core-hole-induced vibrational effects, which are especially strong in the case where the intermediate potential energy surface is dissociative[2].

In the case where the energy of the incoming photon is large enough to eject an electron from the system the KH formula loses its dependence on the energy and polarization direction of the incoming photon[2]. This is because a continuum electron can have any energy and polarization direction. We can then forget about the ejected electron and only consider the charged system containing a core hole, and the final states, which will have an electron missing in the valence. For this system we use the non-resonant KH formula for the emission intensity σ at frequency ω

$$\sigma(\omega') \propto \sum_F \sum_{f_F} \left| \sum_{n_N} \frac{\langle f_F | D'_{FN} | n_N \rangle \langle n_N | D_{NI} | i_I \rangle}{\omega' - E_{n_N f_F} + i\Gamma} \right|^2$$

Eq. (1)

where the upper case letters denote electronic states and lower case vibrational states, with a subscript denoting the electronic state to which they belong. The initial state is denoted by *i*, the intermediate by *n*, the final state by *f*. *D* and *D'* are the dipole operators, $E_{n_N f_F}$ is the transition energy between the states $n_N$ and $f_F$ and *Γ* is the HWHM lifetime broadening parameter.

For systems with very few vibrational degrees of freedom it is possible to compute all electronic potential energy surfaces and their corresponding vibrational states, and directly use Eq. (1) to compute the spectrum. The SCKH formula can be derived as an approximation to Eq. 1, by going over to the time domain and letting the nuclear degrees of freedom move classically, while the electrons are still described quantum mechanically[34], giving the expression

$$\sigma^{SCKH}(\omega') \propto \sum_{traj} \sum_{F} |D_F^{SCKH}(\omega')|^2$$

Eq. (2).

where the first summation (traj) is over dynamical trajectories on the core-hole state and the second sum is over final states $F$, and

$$D_F^{SCKH}(\omega') = \int_0^\infty dt' \, D'_{FN}(t') D_{NI}(0) e^{-i \int_0^{t'} d\tau E_{NF}(\tau)} e^{-\Gamma t'} e^{i\omega' t'}$$

Eq. (3).

In Eqs. (1), (2) and (3) there is an implicit assumption that the Born-Oppenheimer approximation holds, that is, that the potential energy surfaces (PES) are not strongly coupled.

The PESes were computed with density functional theory using the deMon2k code [35]. The standard Perdew–Burke–Ernzerhof (PBE) gradient-corrected exchange-correlation functional [36] was used. For carbon and hydrogen the [4$s$/3$p$/1$d$] valence triple-zeta basis set [37] and the IGLO-II [3$s$/1$p$] basis set[38] were used, respectively. To describe the core-excited oxygen in the central methanol molecule the IGLO-III basis set[38] was used while the remaining oxygens were described with relativistic effective core potentials and [3$s$/3$p$/1$d$] basis sets [39]. The same computational parameters were used for the transition amplitude calculations along the trajectories. The ground state was computed with spin-restricted density-functional theory (DFT) and the energies of the valence-ionized states were approximated by the ground state total energy minus the corresponding orbital energy. The core-ionized state was computed by explicitly removing one (spin alpha) core electron, leaving the system charged, and relaxing the electronic structure in a spin-unrestricted manner [40]. Transition amplitudes were computed at each time step along the trajectories using the ground state orbitals, *i.e.* neglecting relaxation effects both in the core-hole and final valence-hole states which has proven to give a balanced description [41].

**III. Results**

*A. Methanol dimer*

In order to ensure that the SCKH method is applicable for typical structures occurring in the liquid we first make a comparison to the full KH method for a model methanol dimer, H-bonded at a range of distances. In Fig. 1 (upper) we show the spectrum for the optimized dimer computed with the KH method, compared to the averaged spectra weighted by the initial state vibrational distribution. It is apparent that only averaging the spectra without including any dynamics gives too narrow peaks that fail to capture the shape of the spectrum; we note, however, that all peaks in the final spectrum are present also in the spectrum without dynamics, albeit with significant differences in the intensities

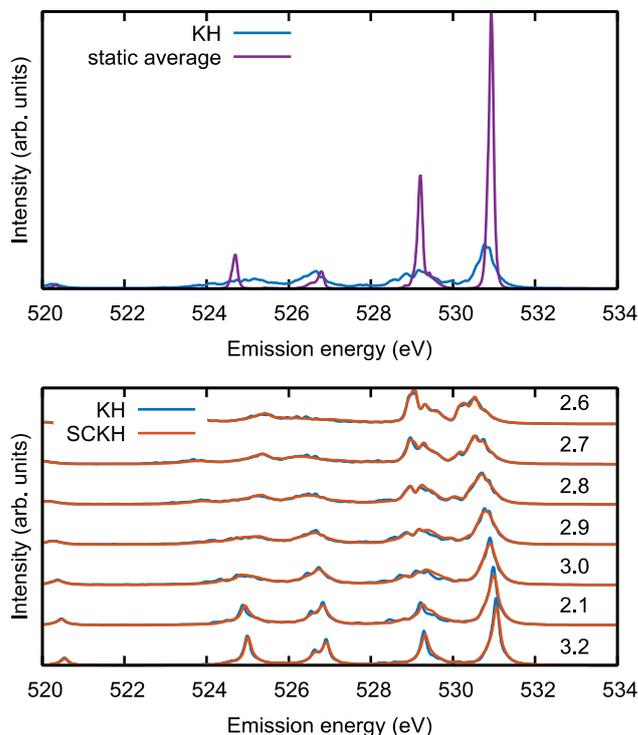

**Figure 1:** (color online) Upper: XES computed for an optimized methanol dimer with O-O distance 2.91 Å. Blue: KH, purple: static average of spectra weighted by the initial state vibrational distribution. Lower: XES for the methanol dimer for different O-O distances (in Ångström), as shown in the plot. Blue: KH, orange: SCKH.

In Fig. 1 (lower) we show computed spectra for methanol dimers with varying oxygen-oxygen distances, from a very short 2.6 Å, compared to the optimized dimer distance of 2.9 Å, to a long 3.2 Å, using the KH and the SCKH methods. The SCKH spectra were computed by summing over 100 trajectories where the initial conditions in terms of O-H distances (10) and momenta (10) were sampled from the ground state vibrational distribution. The trend in the spectra is that for shorter bonding distances the peaks get more spread out while for longer distances they become narrower. Apparently the dynamical effects are more important for short distances, while for the longer ones one approaches the case where averaging spectra from the initial state is a good approximation. This is fully consistent with the results for water dimer [20] and water clusters [10] where the resulting dynamics for non-resonant excitations were found to be along the H-bond direction. For all computed structures, the SCKH method gives almost identical results to the fully quantum mechanical description. This agreement is similar to that observed for a water dimer in ref [20].

To investigate the trends further we look at the potential energy surfaces for the core-ionized state, as shown in Fig. 2. We see that for short O-O bonding distances the PES is dissociative, meaning no minimum close to the ground state equilibrium distance, instead it has its minimum close to the accepting oxygen and the spectrum will thus display large dynamical effects when the molecule is core excited. For longer O-O distances the potential displays more of a double well shape, and for sufficiently long distances, such as the shown 4.0 Å distance, there will be only limited dynamical effects since there is not energy enough to cross the barrier to the other well.

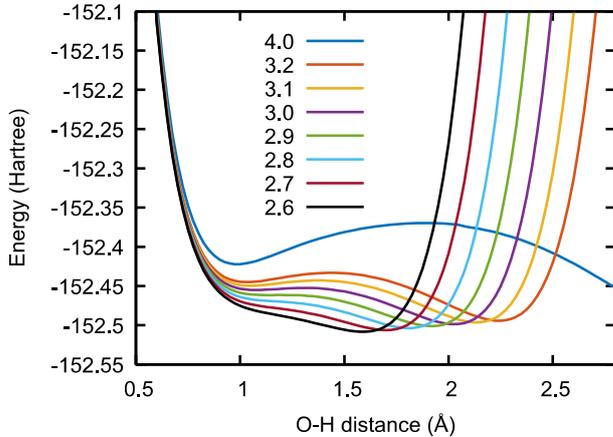

**Figure 2:** (color online) Potential energy surfaces for the core ionized state of methanol dimers with varying O-O distance from 4.0 Å to 2.6 Å, as shown in the key.

In this way we see that the bonding situation in the liquid will be crucial for the dynamical effects on the XES spectra.

*B. Liquid methanol*

For the case of liquid methanol, clusters containing 17 molecules were extracted from a classical molecular dynamics (MD) simulation at room temperature [42]. The central molecule is the one that is to become core-ionized. For each structure, an ensemble of classical *ab initio* MD trajectories were run on the core-ionized state, using sampled quantum initial conditions for the OH vibrational mode of the central methanol molecule. The trajectories were run for 40 fs with a time step of 0.25 fs using the velocity Verlet algorithm with no thermostat. The initial velocities for the other atoms were set to zero. In total we used 40 clusters, with two samples of the initial x-coordinate and four samples of the momentum (two in each direction). Due to some convergence problems some of the trajectories were removed from the set, resulting in a total of 294 trajectories used for the averaging for MeOH and 303 trajectories for MeOD. For each point along the trajectory the energies for the ground state, the core ionized state and the valence ionized states were computed, as well as the transition dipole matrix elements from valence to core orbitals. Spectra were then calculated using the SCKH method, following Eq. (2). In order to have a good comparison for the case of deuterated methanol we made use of the same starting structures as for the normal methanol, only the masses of the hydrogen atoms belonging to the OH-group were changed and the initial velocities were set to correspond to sampling of the ground state vibrational wave function of deuterium.

*1. Core-hole-induced dynamics*

By inspecting the trajectories we could draw some important conclusions about the effect of the core-hole-induced dynamics. We first discuss MeOH. In approximately 75% of the cases the hydrogen visibly bounces back and forth between its originally covalently bonded oxygen and its H-bonded one, as indicated by the one-dimensional model. However, the movements of the other atoms were seen to be very important, especially the bend modes involving the core ionized OH group seem to become excited. Thus, the one-dimensional model, although qualitatively correct in most cases, cannot be expected to quantitatively account for all the relevant effects on the XES spectrum of the core-hole-induced dynamics. Indeed, we checked the one-dimensional model for selected clusters and obtained as good agreement between KH and SCKH as for the dimer, but observed large differences compared to the case where the full dynamics was used. In some of the cases the hydrogen atom leaves to recombine with the other methanol molecule, and after one or two vibrational cycles it returns again. In a few cases a concerted proton transfer was observed, where the transferred proton induces the second methanol molecule to transfer its original hydrogen to a third methanol molecule. Actual dissociation where the hydrogen atom totally leaves its methanol molecule was occasionally observed, but in most cases the hydrogen becomes effectively shared between the two molecules. This group of structures (the ~75% as discussed above) consists of predominantly strongly H-bonded central molecules, as will be discussed later in detail. In about 25% of the cases (mostly non-H-bonded or weakly H-bonded structures) the hydrogen does not have enough energy to overcome the potential barrier and stays close to its original oxygen. Instead, the whole OH group is observed to slowly dissociate from its $CH_3$ fragment. For MeOD we observe almost identical dynamical behavior as for MeOH, with the important difference that the deuterium dynamics is slower due to the larger mass involved.

We can gain a further understanding of the effects of the dynamics on the spectra by looking at $E_{NF}$, the intermediate state energy minus the final state energy as a function of time along the trajectories. The oscillations of $E_{NF}$ enter in Eq. 2 and will give rise to the vibrational broadening effects in the spectrum. Two typical trajectories, for an H-bonded (upper frame) and non-H-bonded methanol molecule (lower frame) are shown in Fig. 3. The transition energies for the H-bonded molecule falls rapidly downwards in energy with time, reaching a minimum at around 7 fs, and then starts slowly oscillating as the hydrogen atom bounces back towards its original methanol molecule. For the non-H-bonded case we instead see a slow monotonous decay of the transition energy. The dynamics should introduce a shift of the peaks in the spectrum mostly to lower energies, and since early times will be weighted more than late times so the short-time behavior will be of most importance we can conclude that the H-bonded species will have more dynamical effects and will have an enhanced intensity at lower energies. We will now make a detailed analysis of the computed spectra for liquid methanol.

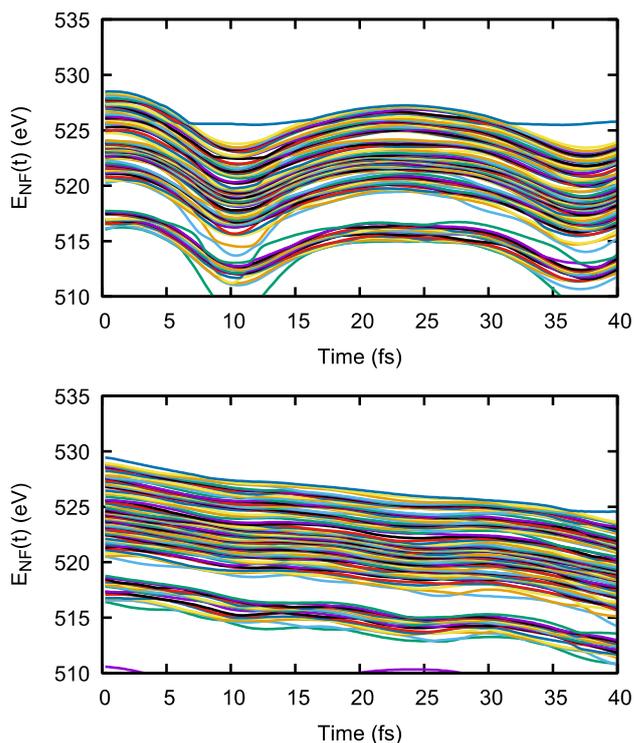

**Figure 3.** Difference between intermediate and final state energies as function of time along the core-hole-induced trajectory. Upper: A typical trajectory for an H-bonded molecule. Lower. A typical trajectory for a weakly or non-H-bonded species.

*2. Analysis of calculated spectra*

In Fig. 4 we show the computed spectra for both MeOH (upper) and MeOD (lower) compared to the experiment of Schreck *et al.* [27]. Since we lack a sufficiently good energy scale the computed spectra are shifted by -2.7 eV to match experiment, furthermore the spectra are area normalized within the shown region. For both cases we observe all the experimental features in approximately the correct position: the split lone pair peak around 526-528 eV, the two lower peaks between 520 and 525 eV, and the weak feature above 515 eV. For MeOH, the computed spectrum is too smeared out compared to the experiment and it is also slightly compressed, with especially the feature at 515 eV occurring at higher energy compared to the experiment. The latter is an expected consequence of the assumption that the neglected relaxation energy in the valence states is a constant and the same for inner and outer valence states; since ionization from inner valence states in reality results in larger relaxation effects this leads to the observed compression of the computed spectrum. The ratio between the two split features matches experiment very well. Looking at MeOD we see that the second split peak has decreased in intensity compared to the first one, which closely matches what is seen in the experiment. The positions of the peaks are essentially unchanged from that of MeOH but are sharper, although they are still a bit too smeared out.

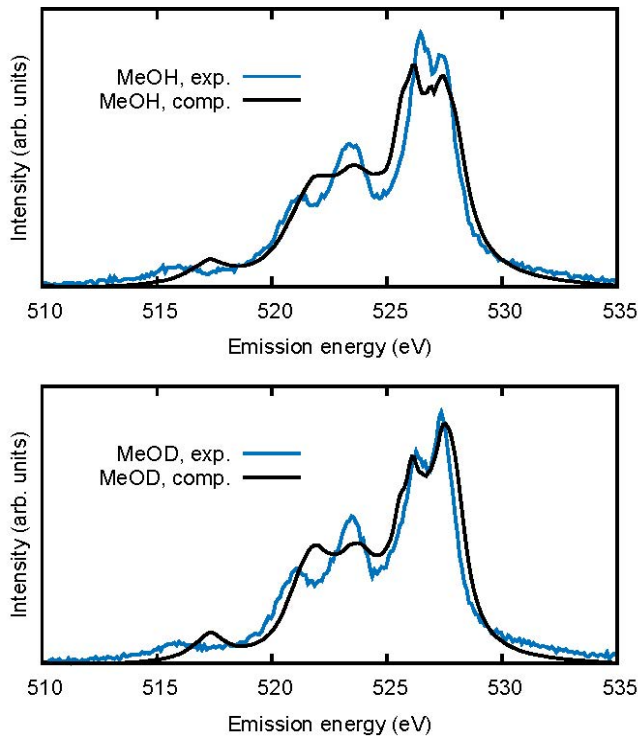

**Figure 4:** SCKH spectra including dynamical effects compared with experiment [27]. Upper: Spectrum computed for normal methanol (black line) and experiment (blue line). Lower: Computed spectrum for deuterated methanol (black line) and experiment (blue line). The area of the spectra is normalized in the shown region.

*3. Dependence on the initial structures*

A central question in the interpretation of XES spectra of liquids is if we can assign different features or contributions to local geometric molecular arrangements[8]. In particular the H-bonding situation is of great interest in the present case, since we know that the potential of a H-bonded hydrogen has a different character than a non-bonded one. We chose to use the cone criterion [43] that considers an H-bond to exist when the oxygen-oxygen distance $r_{OO}$ is smaller than an expression that depends on the maximum distance in Ångström for a totally straight bond, $r_{max}$, and the angle in degrees between the donating oxygen atom and the hydrogen atom:

$$r_{OO} < -0.00044\theta^2 + r_{max}$$

In ref. [43] $r_{max}$ was set to 3.3, here we also look at the more restrictive values of 3.1 and 2.9 in order to have more refined criteria. Our 40 sampled structures were found to have 86 % H-bonded molecules at $r_{max}$=3.3, 73% at $r_{max}$=3.1 and 33 % at $r_{max}$=2.9 while the statistics for the whole MD-dump from which the structures were extracted was 86 %, 72% and 45% respectively. In terms of hydrogen bonds, our sampling is then very good except that we have approximately 30% more H-bonded molecules between 3.1 and 2.9 than in the dump, with the same contribution lacking in the category with the shortest bonds.

In Fig. 5 we show the spectrum obtained as a static average of the different initial structures, without including any dynamics, and its decomposition according to H-bond situation. The peaks are very narrow compared to the experiment.

In the upper frame we see the general behavior that the non-bonded or weakly bonded situations contribute at higher emission energies. This is more clear when zooming in to the two first peaks as shown in the lower frame. Although the correlation between H-bond situation and energy positions is not perfect we see that strongly bonded situations generally contribute at lower energies and weakly bonded at higher energies. This core level shift is up to 0.5 eV in the extreme cases and the total spectrum shows a splitting of about 0.2 eV. We note that, in our description of the electronic structure we describe the ground state and the core hole state well using DFT total energies, however the valence hole energy also includes an orbital energy, making the core level shift of the XES peak more uncertain.

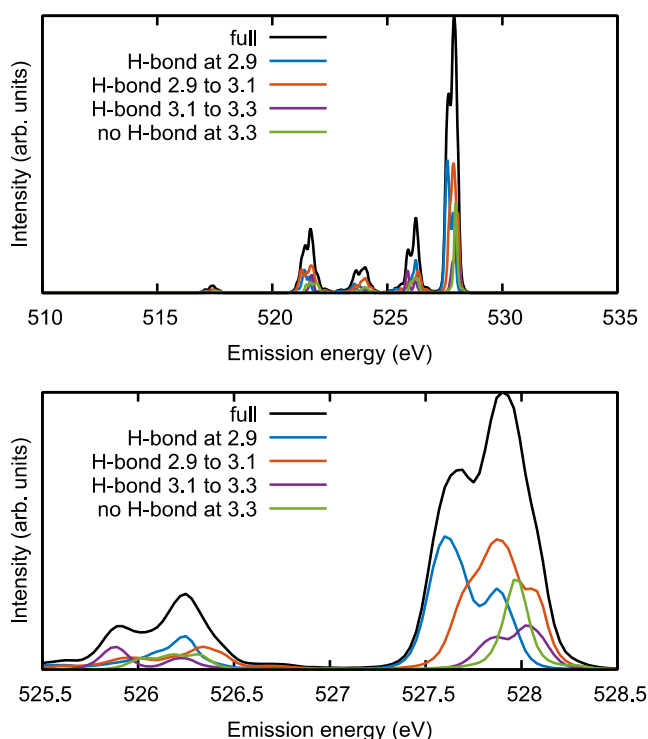

**Figure 5:** Contributions to the static t=0 spectrum from different hydrogen bonding situations. The black line denotes the full spectrum. Blue line: contributions from molecules H-bonded at the r-parameter 2.9 according to the cone criterion, orange line: H-bonds between 2.9 and 3.1, purple line: H-bonds between 3.1 and 3.3, green line: no H-bond at the r-parameter 3.3. The upper frame shows the full spectral range while the lower frame is a blowup in the range of the first two peaks.

Now we include the dynamical effects and do the same comparison, as seen in Fig. 6. In the upper frame the contributions are plotted to scale and we can see that also here the strongly H-bonded situations contribute to the lower peak and the weakly bonded ones more to the higher peak. To better see the shape of the different contributions they are plotted with their area normalized in the lower frame of Fig. 6.

As we already have remarked, the H-bonded species show the type of dynamical behavior in which the hydrogen atom is ejected and becomes shared with the accepting methanol molecule, and now we can connect such behavior with the lower-energy split emission peak. Indeed, by inspection we can see that the structures that display the most dissociative dynamics also have the most strongly peaked contributions at the lower energy range.

Furthermore, for the first split peak the dynamical effects seem to go in the same direction as the core level shift. In other words: structures with strong hydrogen bonds can be expected, on average, to have a downshifted static spectrum as well as strong dynamical effects that shift down the spectrum even further, and for weak or non-existing H-bonds the opposite seems to be true, they will have static spectra to higher energies and less dynamical effects, making them contribute more to the high-energy split peak feature.

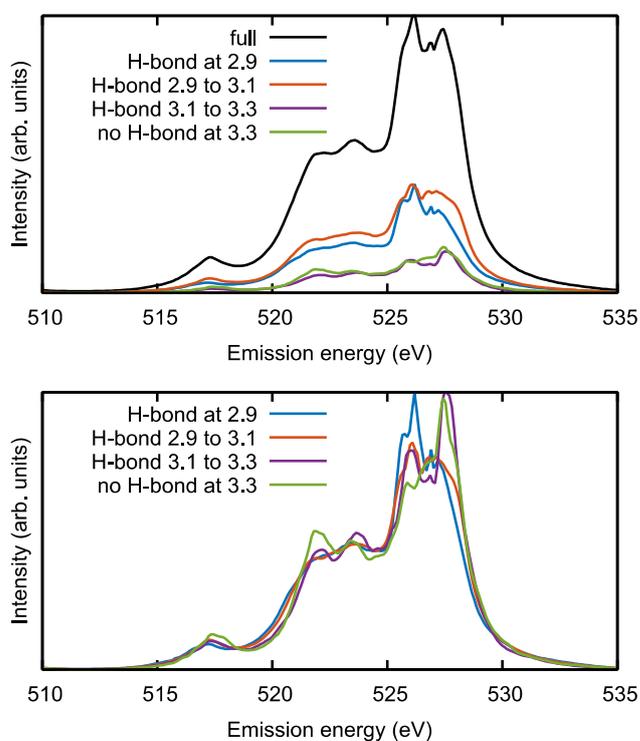

**Figure 6:** Contributions for different hydrogen bonding situations for MeOH including dynamical effects. Upper: the black line is the full spectrum and the other contributions denote various H-bonding situations according to the r-parameter in the cone criterion. Lower: Here the different contributions of the upper frame have each been area normalized. The full spectrum is not shown.

*4. Dependence on the quantum initial conditions*

Another factor that could influence our results is the quantum initial conditions that we use. In Fig. 7 we show the contributions to the MeOH spectrum from the different samplings of position and momentum, where the upper frame shows all four values of the momentum for the short OH distance and the lower frame shows them for the long OH distance. We can immediately observe that the short distance has more of the low split feature, that is more dynamics, and the long distance has more contributions to the higher peak, for all momenta. Our interpretation is that a short OH distance means that the hydrogen atom is up the slope of the potential well for the core-hole (intermediate) state and so gains velocity in the outward direction when it starts to move. For the long value of the OH distance the potential is more flat and not a lot of outward velocity is gained due to the potential. Looking at the different momenta, it is clear that the one of high magnitude pointed outwards induces more dynamical effects in comparison to all others, which are more similar. Nevertheless one can see the expected behavior that more outward momentum promotes more of the lower energy peak.

By dividing up the different hydrogen-bonded situations by quantum initial conditions (not shown) we can see that for the non-H-bonded species there is very little contribution to the low-energy peak, no matter the initial condition. For the H-bonded species where the dynamical effects are important, some initial conditions promote a fast hydrogen dissociation and others make it more slow, with the corresponding effect on the peak shape. Especially molecules with an intermediate H-bond strength can be tipped one way or the other by the initial conditions and thus display the largest effect on the spectrum.

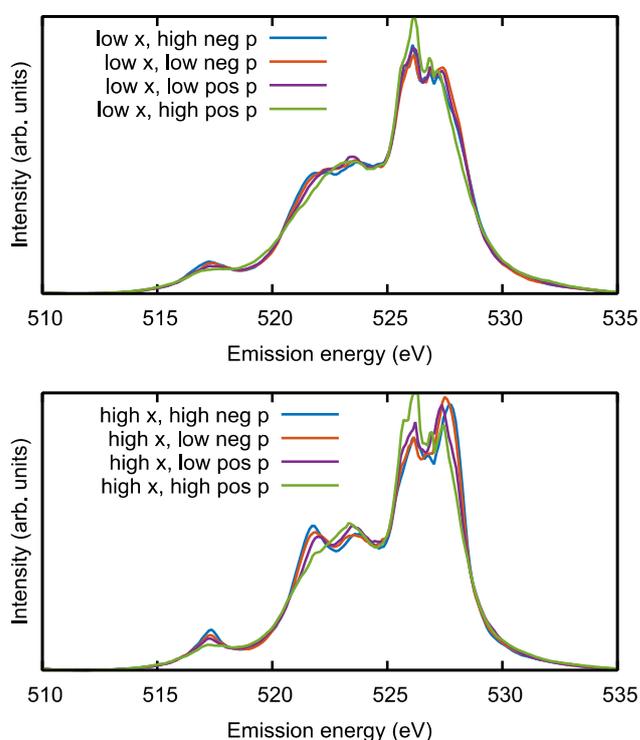

**Figure 7:** Contributions of different sampled quantum initial conditions for MeOH. Upper: all used momentum samples for the short x value. Lower: all momentum samples for the long x-value. Blue: high magnitude of *p* directed inwards to the covalently bonded oxygen, red: low magnitude of *p* directed inwards, purple: low magnitude *p* directed outwards, green: high magnitude *p* directed outwards.

*5. Orbital decomposition*

To further aid us in the interpretation of the spectrum we decompose the contributions coming from different final states, as shown by the coloring of the peaks in Fig. 8. For a single methanol molecule the molecular orbitals can be categorized according to their symmetry; highest in energy we have the $2a''$, then comes the $7a'$, and then almost degenerate $1a''$ and $6a'$, followed by $5a'$ and $4a'$. The lower-lying orbitals will not contribute within the energy range of interest. For the liquid clusters with 17 molecules we make an approximate decomposition by assigning the 17 highest final states to $2a''$, the next 17 to $7a'$, etc. As can be seen in Fig. 8, although there are small contributions that end up outside of the main features, this decomposition gives clearly separate peaks that we will be able to follow and see where they end up when the dynamics is included.

In the upper part of Fig. 8 we show the orbital contributions for the static spectrum. The different contributions form clearly separated peaks that do not overlap. In the middle part of Fig. 8 we show how the orbital contributions look for MeOH, when the dynamics is included. The narrow peaks in the static spectrum have been smeared out to give a much broader spectrum, where we have a general broadening of all contributions that now overlap significantly. The first split peak is seen to originate from a combination of $2a''$, which broadens down to lower energies and covers both peaks of the split experimental peak, and $7a'$, mostly contributing to the lower energy split-peak feature; there are also smaller contributions from the tail of the $1a''/6a'$ orbital distribution. Similarly for MeOD we observe broadening of all contributions, although less than for MeOH, and especially the $2a''$ peak becomes less asymmetric towards lower energy, leading to a much reduced intensity of the lower-lying split feature due to less background to the $7a'$ peak.

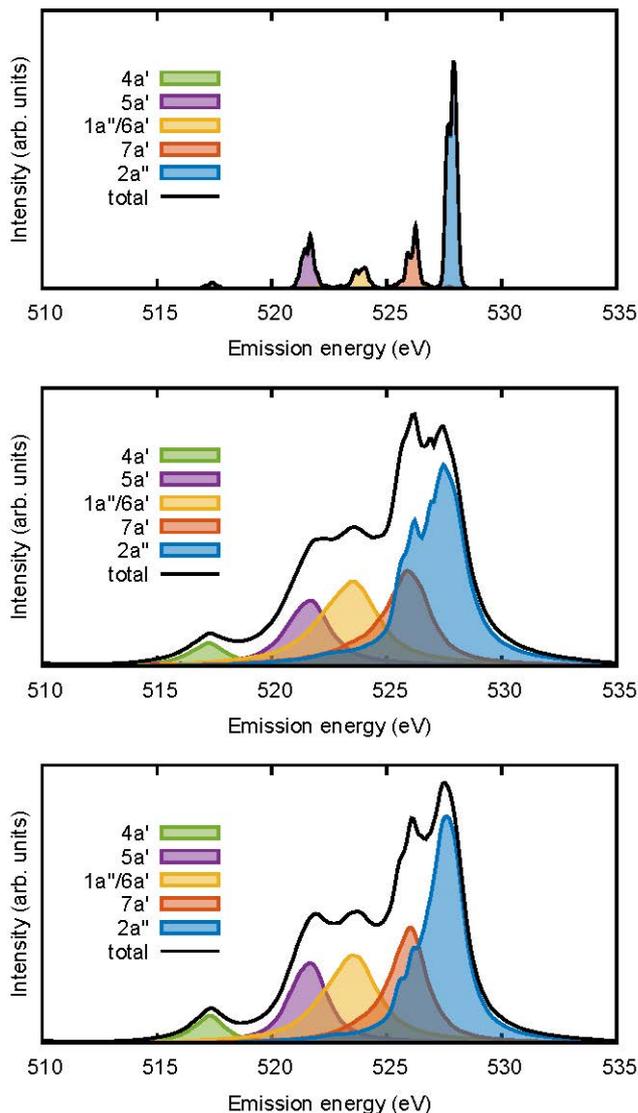

**Figure** 8**:** SCKH spectra (black line) and their orbital contributions shown as fields of different colors. Upper: Statically averaged spectrum at t=0, Middle: SCKH spectrum for MeOH. Lower: SCKH spectrum for MeOD. The orbital assignments go from the lowest emission energy (4*a′*) to the highest (2*a″*).

*6. Time-dependent orbital assignment*

In the adiabatic SCKH method that we use we need to follow each final state in time in order to compute its contribution to the emission spectrum according to Eq. (2). However, as seen in in Fig. 3 the states for a cluster of 17 methanol molecules are closely spaced and often cross each other, and this leads to problems of assigning states between different time steps. The simplest and most naive approach is to assign them according to their energy ordering, but this will not preserve the symmetries when states cross. Ideally, a non-adiabatic formalism should be used that can treat all states on the same footing and compute the transition matrix elements between them. In the present publication we have not attempted such a method, instead we have tried to approximately estimate such effects by sorting the curves between time steps such that we preserve the orbital symmetry. This is done by computing the overlaps of the determinantal wave functions between time steps and order them according to the maximum overlap, and in case of doubts we also take into account the transition dipoles that should match for the symmetries to be the same. Using this procedure we can in most cases see a visual improvement, where we smoothly follow the correct state through a state crossing. In Fig. 9 we show the effects of this reordering on the spectra. For both MeOH (upper frame) and MeOD (lower frame) we observe a sharpening of the features and an enhancement of the high-energy peak in comparison to the unordered spectrum. The agreement with experiment is slightly worsened by the sorting with the higher split peak feature coming up too high, however.

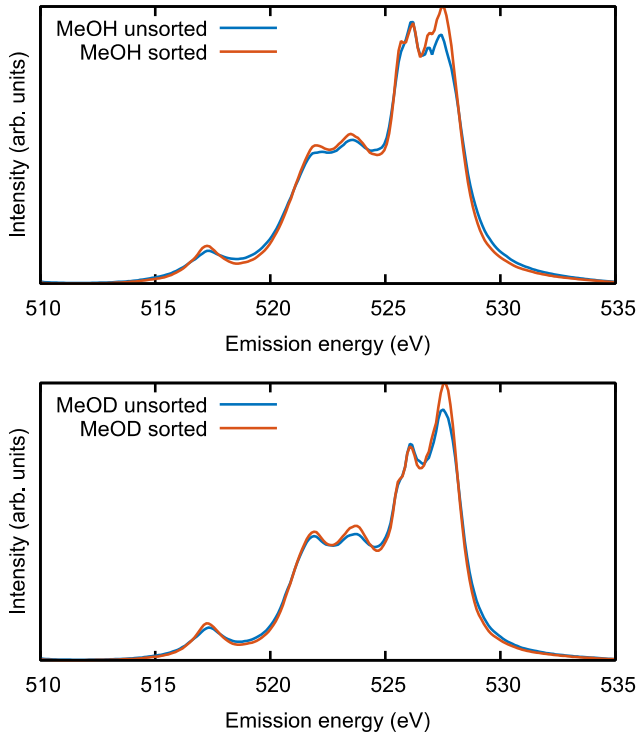

**Figure 9:** Effect of sorting the time-dependent final states. Upper frame: MeOH, lower frame: MeOD. Blue line is the unsorted and the orange line is the sorted.

## IV. Discussion and Conclusions

We apply the SCKH approximation[20, 34] to the case of methanol as one-dimensional model dimer and as liquid. In the first case the restriction to an effectively one-dimensional PES allows to compare the spectrum computed either with a full quantum mechanical description of the core-hole induced vibrational effects or with classical dynamics with initial conditions (positions and velocities) from a sampling of the quantum mechanical distributions of O-H distances and momenta. The excellent agreement provides confidence in the approach and sampling for systems with more degrees of freedom. For the liquid, similar to the case of the dimer, we find that including dynamics only redistributes intensity between the peaks and broadens them, but does not generate new states; this is true even though the PES is formally dissociative with a minimum at the H-bond accepting oxygen.

Including the dynamics and life-time vibrational interference effects using the SCKH approach leads to an excellent agreement between simulation and experiment[27], in particular for the split peak at the highest emission energy where the large isotope effect is accurately reproduced. A decomposition of the split peak in terms of the molecular orbitals reveals the origin of the lower energy feature as a combination of the 2a″ peak broadened to lower energies, providing a background and lifting up the broadened 7a′, and with contributions from the tail of the 1a″/6a′ peak.

In our analysis we have shown that there are several effects that affect the spectral shape of XES on liquid MeOH and MeOD: i) a static core-level shift, ii) dynamical effects, iii) quantum initial conditions and iv) effects of reordering the time-dependent assignment of the states. Of these i) and ii) can be directly related to the local H-bonding structure, where short and strong H-bonds give rise to a static downshift of the emission energies, as well as dynamical effects that also shift the distributions to lower energies. As can be seen from comparing MeOH and MeOD, the dynamical effects are crucial in order to obtain a correct lineshape. The quantum initial conditions iii) are also of importance since they affect how fast the dissociative dynamics will occur. For the non-H-bonded species this effect is small since no hydrogen dissociation takes place for any initial conditions. Finally, there are also some effects due to the ordering in time of the states iv) that affects the peak ratio in the split double peak.

Although the match to experiment is impressive the computed spectra are still slightly too broad, and the positions of the peaks do not match perfectly, the computational spectrum being compressed as compared with the experimental one. This is most likely connected with our DFT description of the valence-excited states, where orbital energies enter. Indeed, we have to shift our spectra by -2.7 eV for the comparison to experiment, and it could be that an individual shift of the spectra should be done before summing which could lead to more well-defined features. To get a better energy scale, TDDFT using recently developed range-separated functionals[44, 45] or Hedin's GW approximation would surely lead to some improvement.

We conclude that including the life-time vibrational interference effects[30] is very important to obtain the proper shape of the spectrum and relative intensities between spectral features. This can be reliably done through the SCKH method using classical dynamics trajectories for the core-hole state with a proper sampling of the relevant quantum probability distributions. The excellent agreement with experiment gives us confidence that our methodology captures the relevant effects of XES for hydrogen bonded liquids and indeed will be suitable in further investigations, not least for the challenging case of liquid water.

**Acknowledgements**
Computational resources provided by the Swedish National Infrastructure for Computing (SNIC) at the HP2CN and PDC centers are gratefully acknowledged. We thank M. O. Hakala, and S. Lehtola for providing structures from their molecular dynamics simulations of liquid methanol, and Simon Schreck for sharing the experimental spectra. OT acknowledges the support by JSPS KAKENHI (No.15K04755).

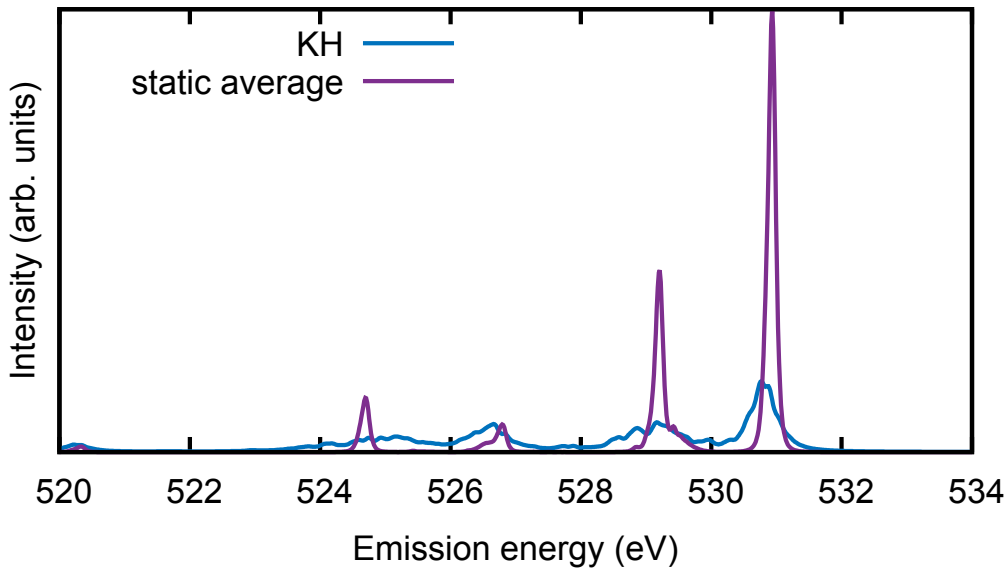
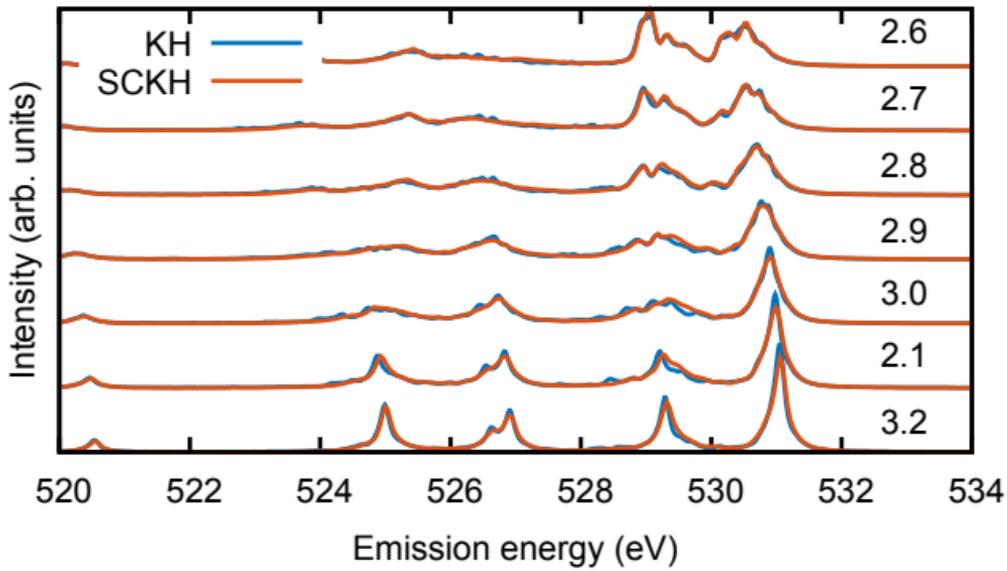

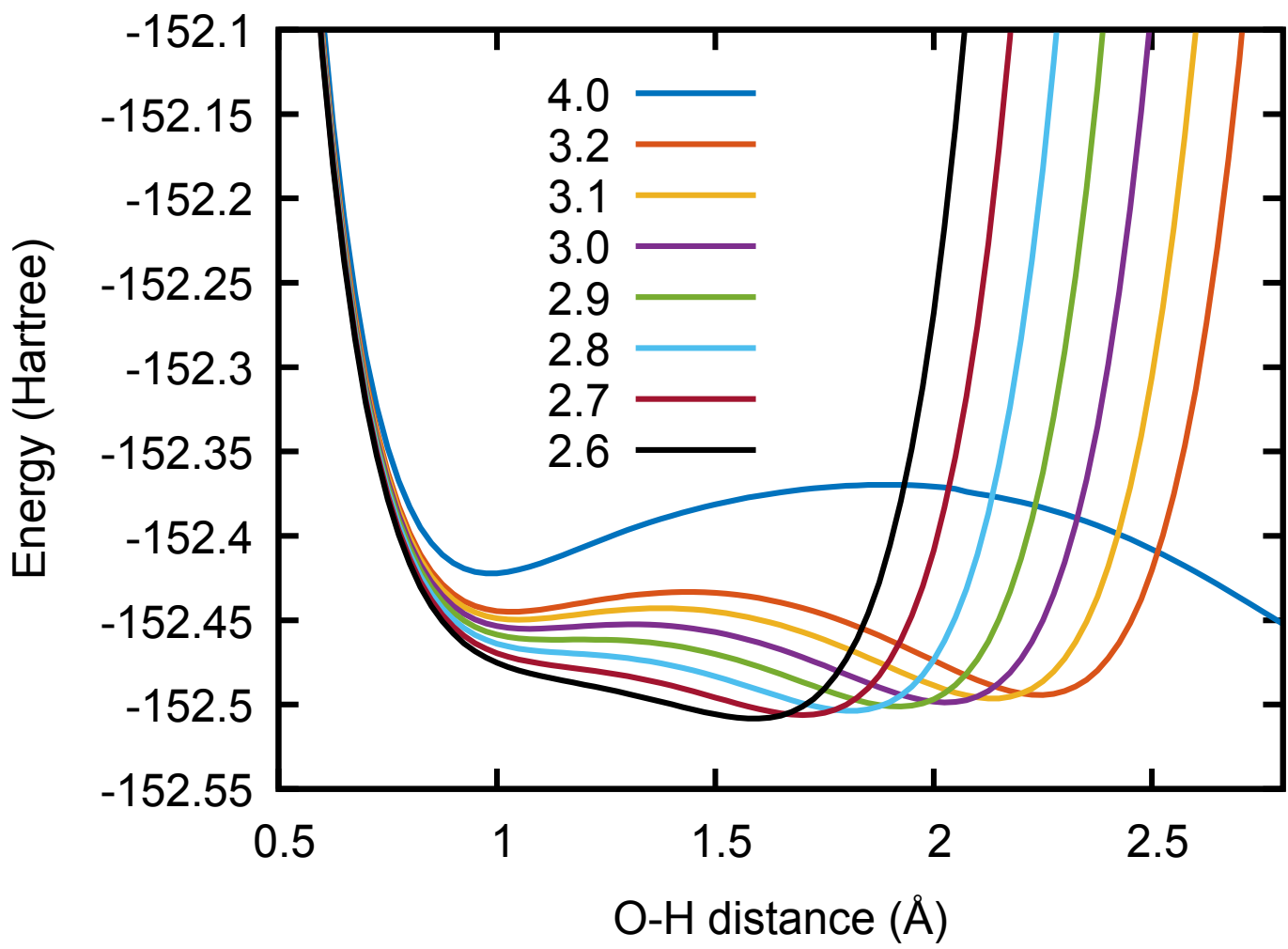

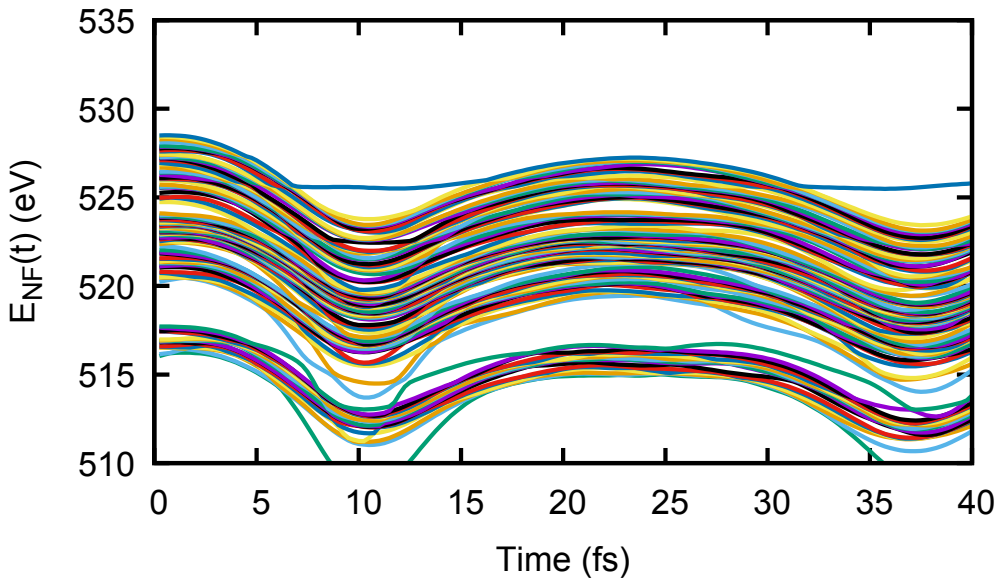
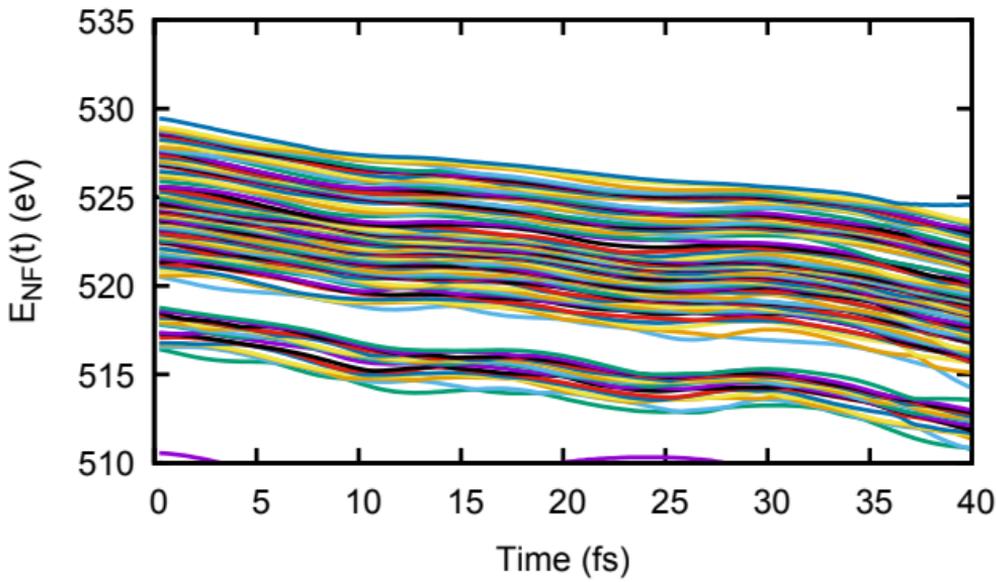

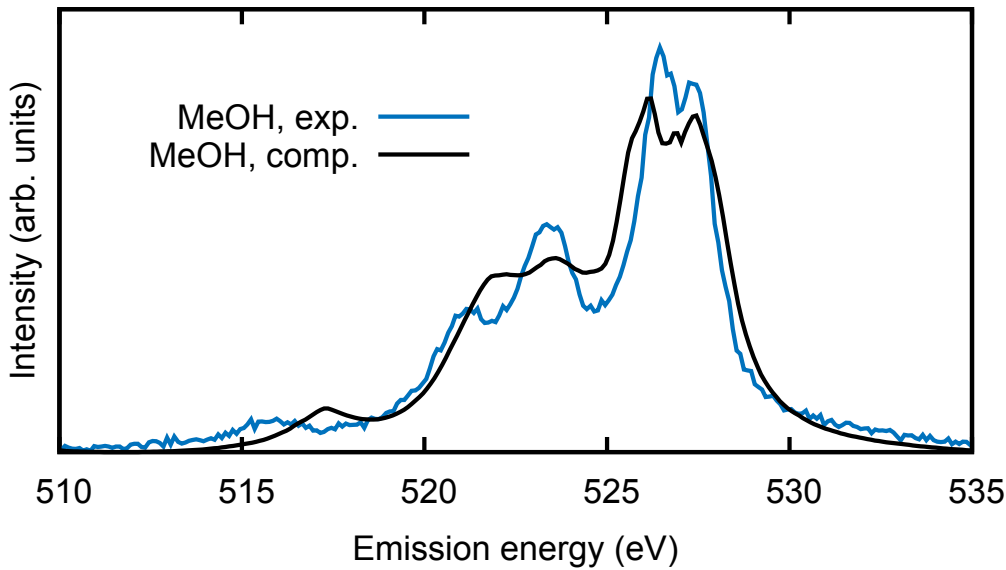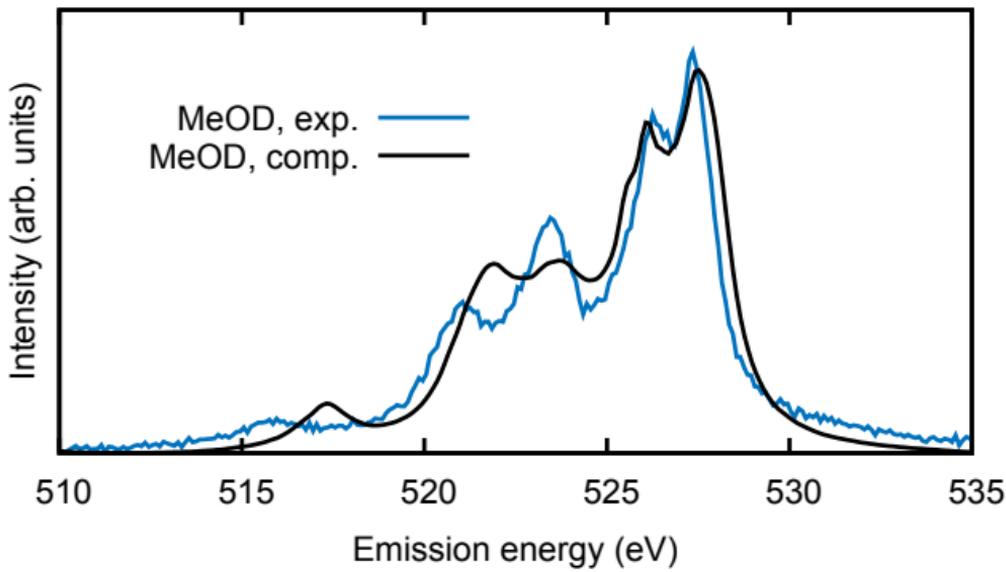

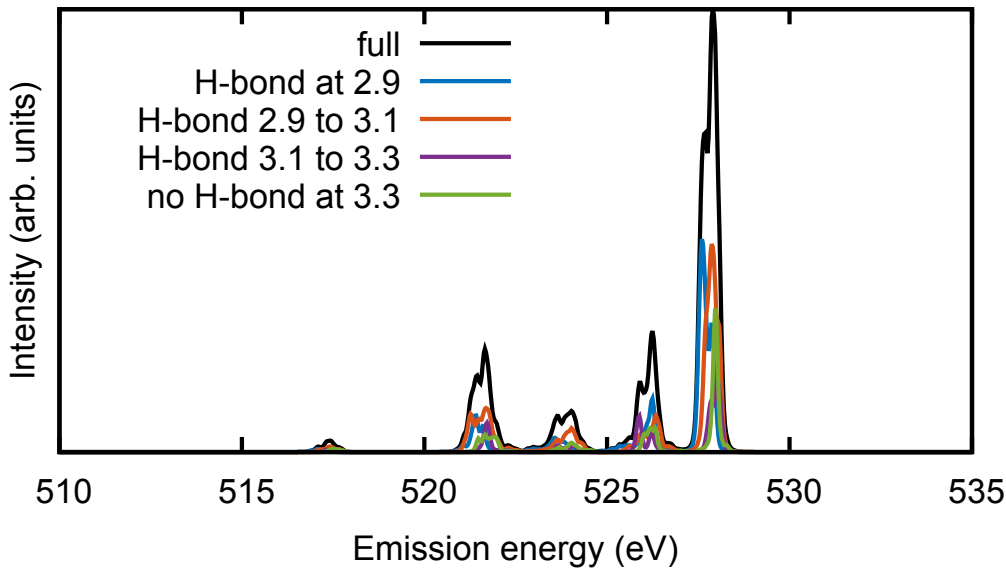
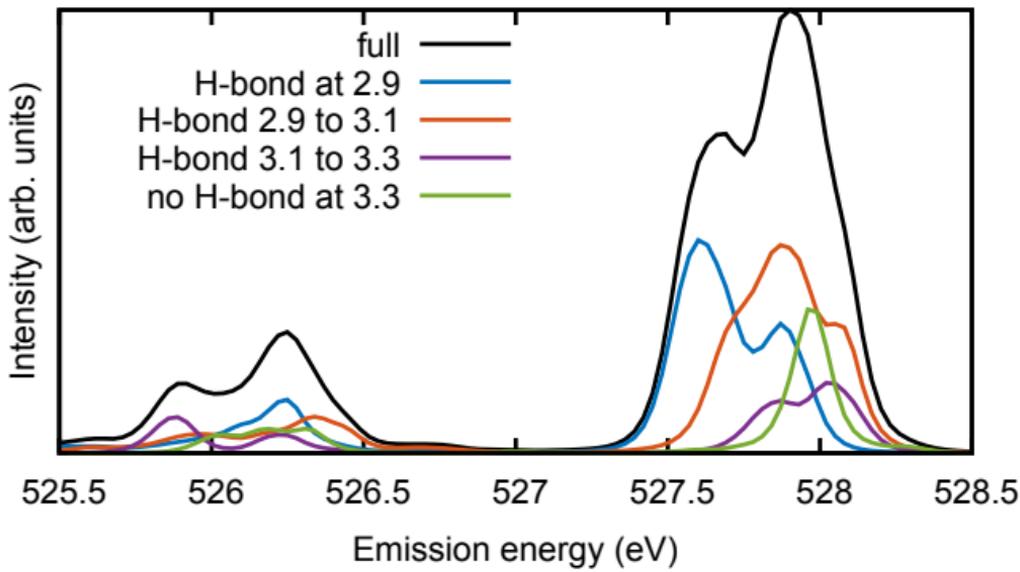

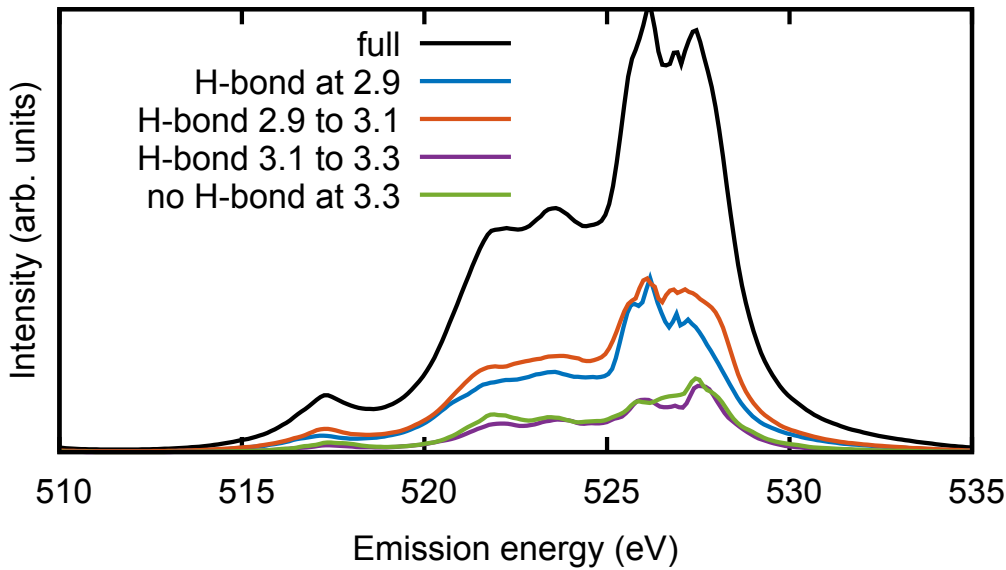
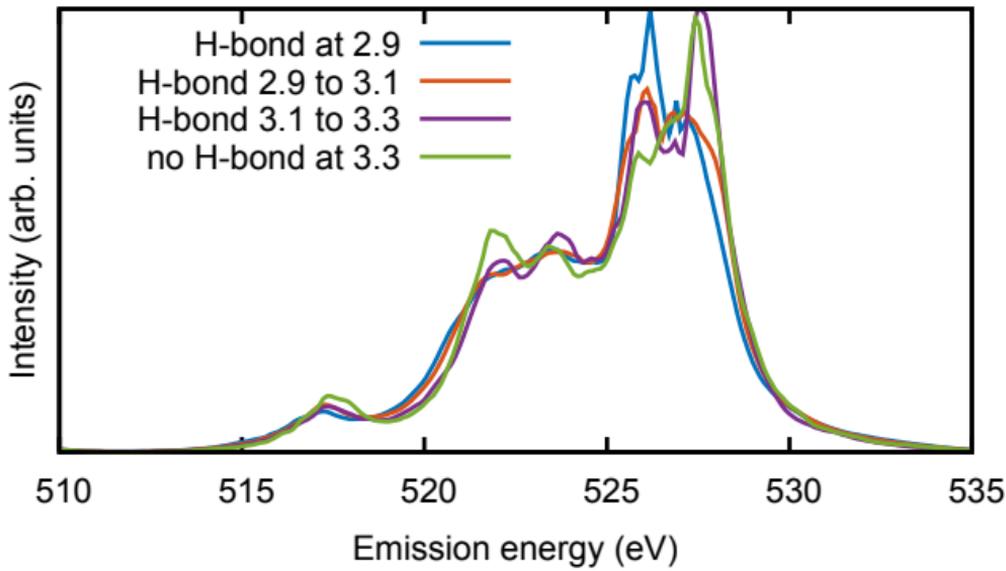

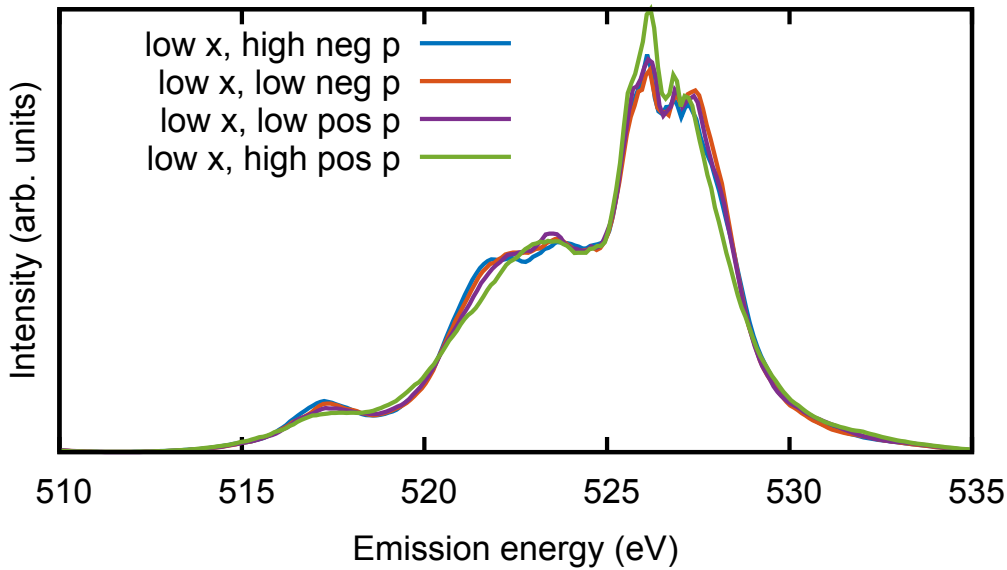
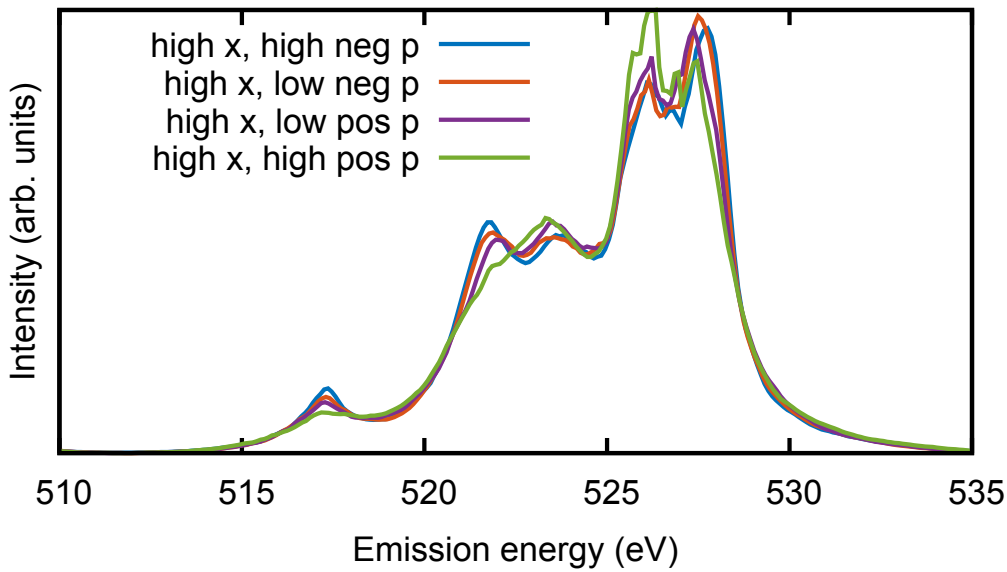

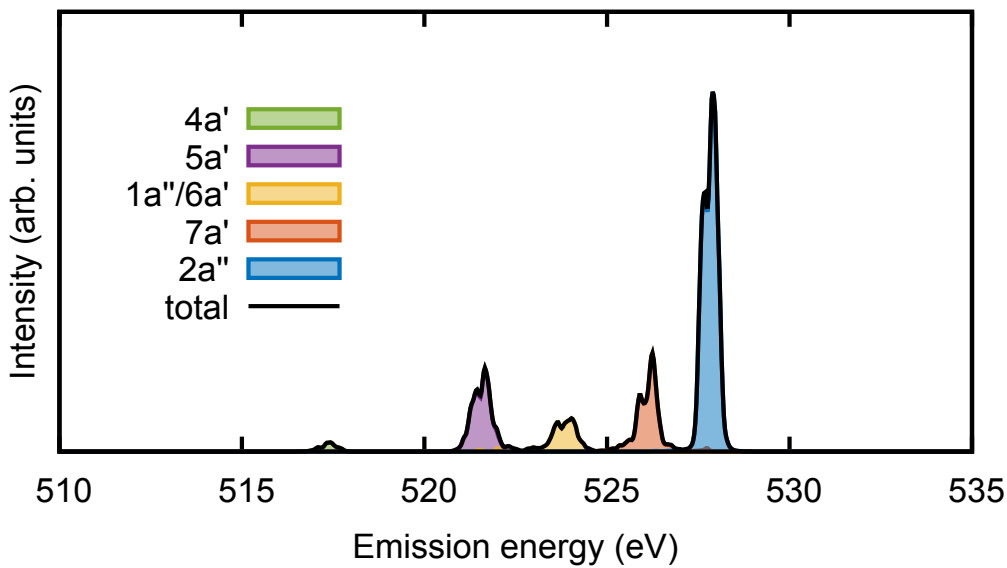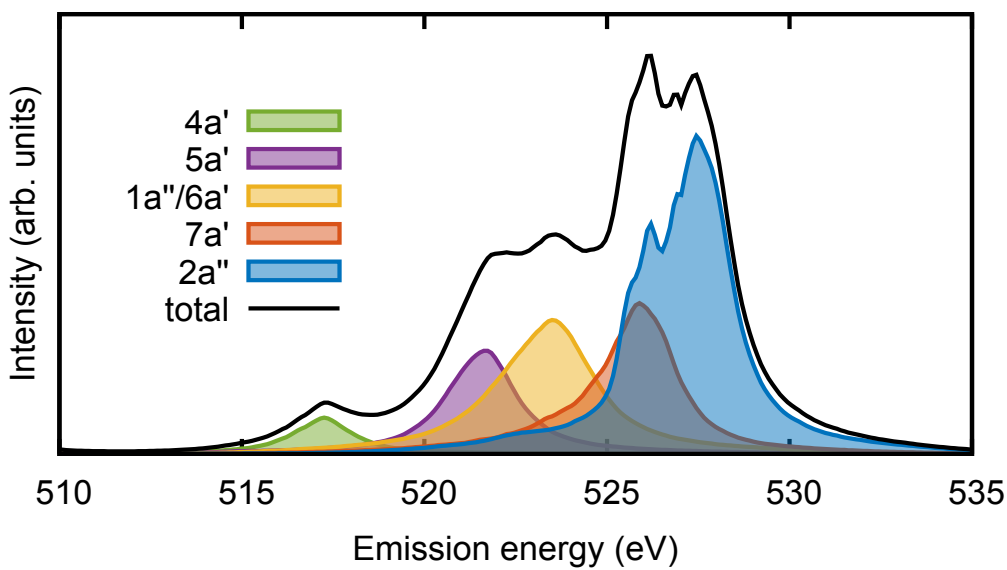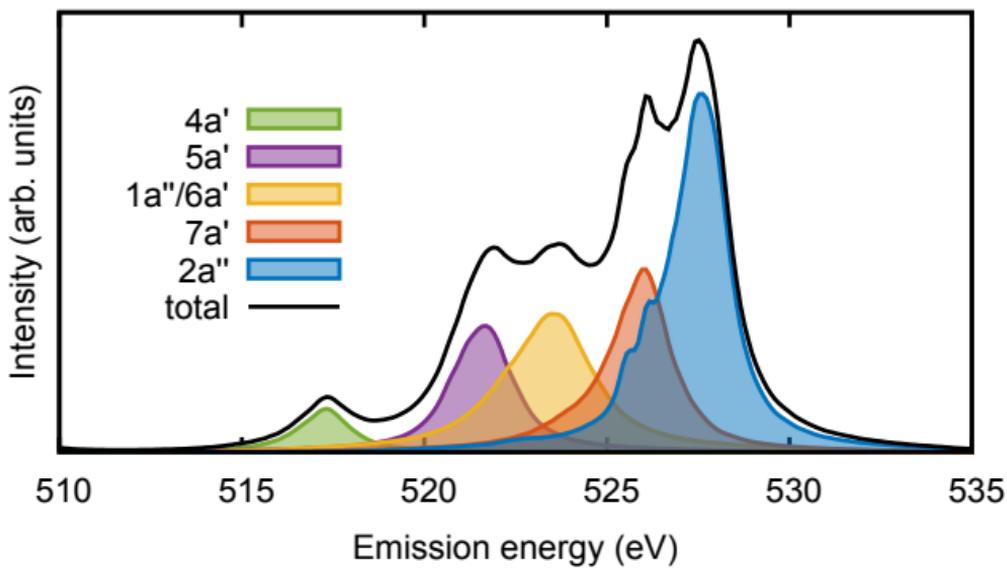

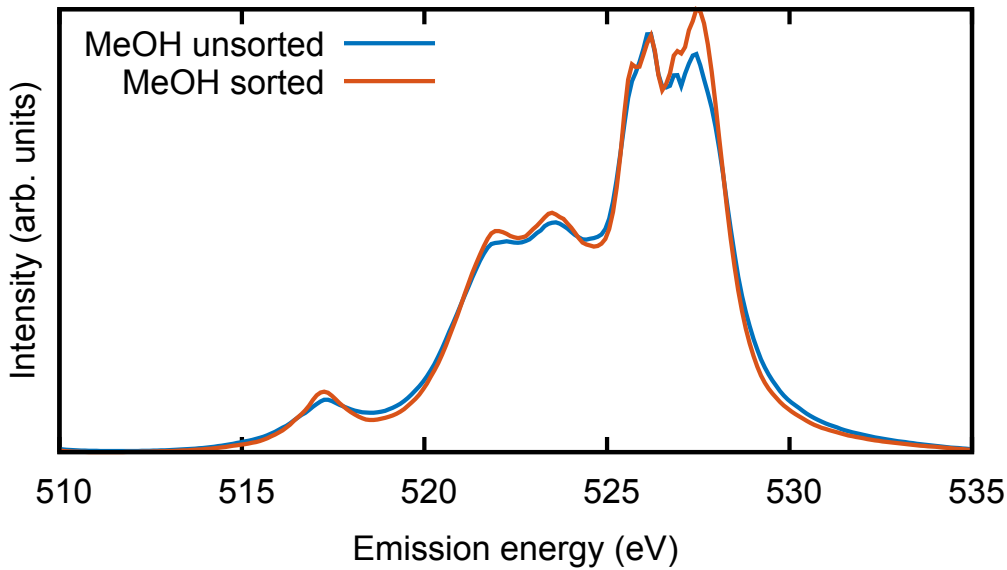
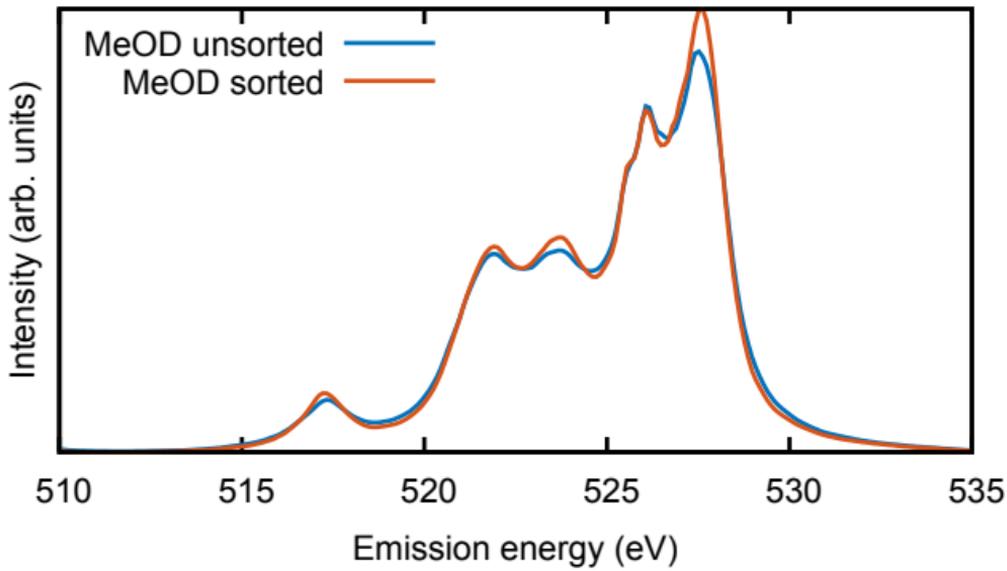